\documentclass{tlp-hacked}
\pdfoutput=1

\usepackage{stmaryrd}
\usepackage{graphicx}
\usepackage[x11names]{xcolor}
\usepackage{wrapfig}
\renewcommand{\arraystretch}{1.2}

\newcounter{mnotei} 
\newcommand{\mnote}[1]{%
{\scriptsize\textsf{\textcolor{blue}{$^{[\themnotei]}$}}}%
\marginpar{\scriptsize\textsf{\textcolor{red}{n.\themnotei: #1}}}%
\stepcounter{mnotei} } 
\renewcommand{\mnote}[1]{}

\usepackage[T1]{fontenc}

\usepackage{listings}
\newcommand*{\Comment}[1]{\hfill\makebox[6.0cm][l]{\rm\it \color{red!70!black}#1}}%
\newcommand*{\Var}[1]{{\tt\textup{#1}}}%
\newcommand*{\VarC}[1]{{\tt\color{red!70!black}\textup{#1}}}%

\usepackage{hyperref}
\usepackage[capitalise]{cleveref} %
\hypersetup{
  colorlinks=true,%
  citecolor=red!80!black,%
  filecolor=red!80!black,%
  linkcolor=red!80!black,%
  urlcolor=blue!50!black
}

\usepackage{tikz}
\usetikzlibrary{patterns}

\usepackage{pgfplots}
\usepackage{filecontents} %

\newcommand{\papertitle}{Lightweight compilation of (C)LP
  to JavaScript
}

\newcommand\paperthanks{The research leading to these results has
  received funding from the Madrid Regional Government under CM
  project P2009/TIC/1465 (PROMETIDOS), and from the Spanish Ministry
  of Economy and Competitiveness under project TIN-2008-05624 {\em
    DOVES}.  The research by R\'emy Haemmerl\'e has also been
  supported by PICD, the Programme for Attracting Talent / young PHDs
  of the Montegancedo Campus of International Excellence.}

\title[\papertitle]{\papertitle\textsuperscript{\thanks{\paperthanks}}}

\author[J.\ F.\ Morales, R.\ Haemmerl\'{e}, M.\ Carro,
  and M.\ V.\ Hermenegildo]{
    Jose F. Morales$^{1}$, R\'{e}my Haemmerl\'{e}$^{2}$, \\
    Manuel Carro$^{1,2}$, and Manuel V. Hermenegildo$^{1,2}$\\
    $^{1}$ IMDEA Software Institute, Madrid (Spain)\\
    $^{2}$ School of Computer Science, Technical  University of  Madrid (UPM), (Spain)
\\
    \email{\{josef.morales,manuel.carro,manuel.hermenegildo\}@imdea.org}\\
    \email{remy@clip.dia.fi.upm.es, \{mcarro,herme\}@fi.upm.es}
}

\newcommand{\compressection}{\vspace{-1em}}
\newcommand{\compressfigure}{\vspace{-1em}}

\date{}

\begin{document}

\label{firstpage}

\pagerange{\pageref{firstpage}--\pageref{lastpage}}
\volume{\textbf{10} (3):}
\jdate{September 2012}
\setcounter{page}{1}
\pubyear{2012}

\maketitle

\begin{abstract}
  We present and evaluate a compiler from Prolog (and extensions) 
  to JavaScript which
  makes it possible to use (constraint) logic programming to develop 
  the client side of web applications while being compliant with 
  current industry standards. %
  Targeting JavaScript makes (C)LP programs executable in virtually
  every modern computing device with no additional software
  requirements from the point of view of the user. In turn, the use of
  a very high-level language facilitates the development of
  high-quality, complex software.
  The compiler is a back end of the Ciao system and supports most of
  its features, including its module system and its rich language
  extension mechanism based on \emph{packages}.
  We present an overview of the compilation process and a detailed
  description of the run-time system, including the support for
  modular compilation into separate JavaScript code.
  We demonstrate the maturity of the compiler by testing it with
  complex code such as a CLP(FD) library written in Prolog with
  attributed variables. Finally, we validate our proposal by measuring
  the performance of some LP and CLP(FD) benchmarks running on top of
  major JavaScript engines.
\end{abstract}


\compressection
\section{Introduction}

The Web has evolved from a network of hypertext documents into
one of the most widely used OS-neutral environments for running rich
applications  ---the so-called Web-2.0---, 
where computations are carried both locally at the browser and
remotely on a server.
A key factor in the success of the Web has been the development of
\emph{open} standards backed up by mature implementations.
One of these is JavaScript~\cite{ecma:es5}, which was initially
designed as a simple dynamic language embedded in HTML documents in
order to offer basic dynamic content.  Factors such as openness,
simplicity, flexibility, full browser integration, and attention to
the security and privacy concerns that naturally arise in the
execution of untrusted code, have helped the language gain very
significant popularity despite its initial low efficiency.
Performance was initially not competitive with %
plug-in based technology like Java-based
\emph{applets}~\cite{ptu:java-jvm}, but  current JavaScript engines in
major browsers use JIT compilation to  optimize the \emph{hot spots}
in the program using type or trace
information~\cite{google:v8/chrome,Gal:2009:TJT:1542476.1542528}.
While still not optimal for computationally intensive tasks,
performance is good enough in many cases, specially those requiring
mostly just graphical user interaction. 
The language has also raised significant interest in the research
community, as witnessed for example by recent work studying the
formalization of the full core language~\cite{MMT-APLAS-TR08}.
Overall it is enabling a disruptive paradigm shift that is
gradually replacing OS-dependent application development with fully
portable Web applications which can run in a variety of devices.
While all this represents significant advances in the technology for
developing Web applications, it is suboptimal to rely on a single
language to solve all problems.
While the whole spectrum of programming languages is normally
available on the server side, server-side execution is not always
appropriate: for example, the client side may not be allowed to
transmit sensitive data %
outside the client, 
and there are always
constraints on network capacity or usage (e.g., local search on a
large set of personal data).%
\mnote{MCL: send queries}
Also, server-side execution requires
computing power and storage on the server dedicated hardware, which
can have an unacceptable cost and/or be a bottleneck for large numbers
of clients.%
\mnote{MCL: huge data centers already running.}
\mnote{MCL: I think that it is not necessary to defend client-side
  computations: they are already here.}

While large applications have been written directly in JavaScript
(despite the lack of analysis tools or a module system), targeting it
as a back end language for cross-compilation is nowadays a popular
option\footnote{See
  \href{https://github.com/jashkenas/coffee-script/wiki/List-of-languages-that-compile-to-JS}{\scriptsize
    \tt
    https://github.com/jashkenas/coffee-script/wiki/List-of-languages-that-compile-to-JS}.} 
in order to execute existing code in web browsers
(since manual rewriting is costly for large projects)
or to use libraries and features available
in other languages.

At the same time, 
there has been interest and significant activity almost since the
start of the Web in programming web applications in Prolog and other 
(constraint) logic programming dialects 
and/or using the Web as a portable graphical interface for
(C)LP programs (including, e.g., complex tools such as analyzers or
theorem provers). \mnote{cite seqprover?}
The major Prolog implementations have focused to date on server-side
execution.  One of the first popular frameworks for developing Web
applications in Prolog is PiLLoW~\cite{pillow-tplp},
where a server running Prolog code communicates with browsers using
HTTP server mechanisms (CGIs) or the HTTP protocol. This same approach
was also taken and extended by SWI-Prolog~\cite{wielemaker:tplp2008}.
\mnote{MCL: I think the server-side discussion is distracting. We
  already said that the server-side can use any language.}
For client-side execution, most systems have targeted Java or the Java
VM. Some of the most notable systems, Jinni and more recently Lean
Prolog, are derived from 
BinProlog~\cite{DBLP:journals/corr/abs-1102-1178}. 
However, in most cases such
systems are developed from scratch and present at least moderate
incompatibilities with server-side systems. Moreover, as technology
shifts from Java towards JavaScript as client-side language, those
systems may suffer from obsolesce. %
There has been one attempt that we are aware of at implementing Prolog
in JavaScript, JScriptLog (\url{http://jlogic.sourceforge.net/}), but
it is an interpreter and is meant to be just a demonstrator, supporting
only a subset of Prolog. 

Our ambitious objective is to enable client-side execution of
\emph{full-fledged (C)LP programs} by means of their
\emph{compilation} to JavaScript,
i.e., to support essentially the full language available on the server
side. Our starting point is the Ciao
system~\cite{hermenegildo11:ciao-design-tplp-short}, which implements a
multi-paradigm Prolog dialect with numerous extensions through a 
sophisticated module and program expansion system
(\emph{packages}). Such packages facilitate syntactic and semantic
language extensions, all of which are also to be supported in our
approach. The module system also offers a precise distinction between 
static and dynamic parts, which is quite useful in the translation
process. Other approaches often put emphasis on the feasibility of the
translation or on performance on small programs, while ours is
focused on completeness and integration: 
\begin{itemize}
\item We share the language front end and implement the translation by
  redefining the (complex) last compilation phases of an existing system. 
  In return we support a full module system including
  packages, as well as the existing analysis and program
  transformation tools.
\item We provide %
  a minimal but scalable runtime system (including built-ins) and a
  compilation scheme based on the
  WAM~\cite{hassan-wamtutorial,Warren83} that can be progressively
  extended and enhanced as required.
\item We offer both high-level and low-level foreign language
  interfaces with Java\-Script to simplify the tasks of writing libraries 
  and integrating with existing code.
\end{itemize}
This allows us to read and compile (mostly) unmodified Prolog programs
(as well as all the Ciao extensions such as different flavors of (C)LP
or functional notation and higher-order, to name a few), run real
benchmarks, and, in summary, be able to develop full applications,
where interaction with JavaScript or HTML is performed via Prolog
libraries and client-side execution in the browser does not require
manual recoding.  To the extent of our knowledge, ours is the first
approach and full implementation which can achieve these
goals. \mnote{clarify contributions further?}

The paper is organized as follows. In \cref{sec:js-nutshell} we
provide an overview of the JavaScript language and introduce our
solution for generating code in a modular way. In \cref{sec:compiler}
we describe the cross-compilation process. In \cref{sec:foreign} we
show the language interface with JavaScript. We present experimental
results in \cref{sec:experimental}. Finally, \cref{sec:conclusions}
presents our conclusions.%
\mnote{MCL: paragraph candidate to vanish}

\compressection
\section{Making JavaScript a target for modular compilation}

\label{sec:js-nutshell}

JavaScript is a simple, lexically scoped, imperative language.
Its syntax is close to that of C or Java but internally 
it is closer to Scheme or Self.
Data objects can be native (numbers, strings, booleans, etc.),
records (mutable maps from primitive data to values), or closures
(anonymous functions).  \mnote{ Remy: I put "objects" at the end of
  primitive. In a enumeration all items should be same nature: either
  all adjectives or all nouns; Jose: OK. I used 'primitive' as a noun
  here. Which sounds better?}
Records contain a distinguished field called \texttt{prototype} that
is the basis for object-oriented programming in JavaScript. When a
field is not found in a record, it is searched recursively 
following the prototype field. This allows records to 
share %
fields (e.g., to implement a class with methods shared by all its
instances). Functions act also as object constructors and are
records themselves, with a special \texttt{prototype}
field.  Given a function \texttt{ctor}, \texttt{new ctor(args)} creates
a new empty object whose prototype will be \texttt{ctor.prototype},
and then executes the function \texttt{ctor} with \texttt{this} bound
to the object. Prototypes may form a
chain, which is useful for implementing inheritance.   The
internal prototype field of a record cannot be accessed directly and the
only valid operation on it is  \texttt{x instanceof C}, which is
true if \texttt{C} is found in the  \emph{prototype chain} of
\texttt{x}.

One of the main drawbacks of JavaScript for developing scalable and
reusable code is the \emph{lack of proper namespaces or module
  system}, where all symbols apparently live in a single common global
namespace.
For that reason, some proposals for adding modularity to JavaScript
programs exist (e.g., Prototype, CommonJS). However, we found them
either too complex or not complete enough for our purposes.
Nevertheless, closures, the scoping rules, and records can be used to
manually achieve effective symbol hiding. We use this mechanism to
implement the necessary symbol tables for encoding \emph{modular}
Prolog programs, as described below.

\vspace{.5ex}
\noindent\textbf{Runtime for Symbol Tables.}~~
Inspired by the implementation of Prolog predicate tables, we defined
a thin runtime layer (\cref{fig:modrt-code}) to implement a symbol
table which associates symbol names (strings) to their
definitions. 
For each symbol we also store associated information,
like \emph{export} tables, used to implement modules.
Additionally, we allow the definition
\mnote{MCL: what does ``definition'' mean here? Need something more specific.}
of symbols associated with
JavaScript classes (coordinating initialization of base classes and
prototype chains).  We will use them to implement the data type hierarchy
for terms and to avoid \emph{tagging}.
We will later (\cref{sec:compiler}) populate tables with actual
definitions (modules, predicates, functors, JavaScript closures, etc.).

\begin{figure}[t]
  \centering\small
\hrule
\begin{lstlisting}[language=c,morekeywords=function,numbers=left]
var UNDEFINED=0, NOT_READY=1, PREPARING=2, READY=3;
@\label{lst:ctor}@function @\$@s(name) { @\Comment{// Symbol constructor}@
    this.name=name; this.exports={}; this.status=UNDEFINED; 
    this.nested={}; this.link=null; this.mlink=null;
}
@\label{lst:def}@@\$@s.prototype.def=function(name, def) { @\Comment{// Define a symbol}@
    var m=this.query(name);
    m.status=NOT_READY;                       @\Comment{// mark the symbol as not ready}@
    def(m); return m;
}
@\label{lst:query}@@\$@s.prototype.query=function(name) { @\Comment{// Query a (sub)symbol}@
    var m=this.nested[name];
    if (m === undefined) { m=new @\$@s(name); this.nested[name]=m; }
    return m;
}
@\label{lst:prepare}@@\$@s.prototype.prepare=function() { @\Comment{// Prepare the symbol}@
    if (this.status !== NOT_READY) return this;
    this.status=PREPARING;               @\Comment{// preparing the symbol (not ready)}@
    if (this.ctor !== undefined) {       @\Comment{// the symbol defines a class}@
        if (this.base !== null) {
            this.base.prepare(); @\Comment{// prepare base symbol, \Var{status} is \Var{READY}}@
            @\$@extends(this.ctor, this.base.ctor); @\Comment{// setup prototype chain}@
        }
        if (this.mlink !== null) this.mlink(this.ctor); // instance methods
    }
    if (this.link !== null) this.link(); @\Comment{// link local from imported symbols}@
    this.status=READY;                           @\Comment{// mark the symbol as ready}@
    // prepare nested symbols
    for (var k in this.nested) if (this.nested[k]) this.nested[k].prepare();
    return this;
}
@\label{lst:extends}@function @\$@extends(c, base) { @\Comment{// (auxiliary for subclassing)}@
    // copy class methods from @\VarC{base}@ to @\VarC{c}@
    for (var k in base) if (base.hasOwnProperty(k)) c[k]=base[k];
    // ensure that the object @\VarC{c.prototype}@ has the prototype @\VarC{base.prototype}@
    function ctor() {}; ctor.prototype=base.prototype; c.prototype=new ctor;
    c.prototype.constructor=c;
}
\end{lstlisting}
\hrule
 \vspace{-1ex}
  \caption{Minimal runtime code for modular symbol tables in JavaScript.}
  \label{fig:modrt-code}
 \vspace{-2.5ex}
\end{figure}

\vspace{0.5ex}
\noindent\textbf{Symbols.}~~
\cref{fig:modrt-code} illustrates our approach for representing
symbols as JavaScript objects. The most important components are  the
\texttt{status}, which stores the initialization state of the symbol,
an \texttt{exports} table which associates names (strings) with
values, and a \texttt{nested} table for associated nested symbols.
Symbols are initially created (constructor in \cref{lst:ctor}) with
an \texttt{UNDEFINED} status.
Initially we create a single \emph{root} symbol (named \texttt{\$r})
in the global scope.
The \lstinline!$m$.query($n$)! method (\cref{lst:query}) obtains the
symbol associated with the name $n$ in the nested table of $m$. If it
does not exist, it is created. Symbol objects behave as pointers or
references to definitions.
Keeping track of nested symbols will  later be useful to ensure correct
initialization, as well as providing a simple way to store tables for
modular programs (e.g., \texttt{\$r.query("lists").query("append/3")},
assuming that the symbol is associated with the predicate
\texttt{lists:append/3}).

\vspace{0.5ex}
\noindent\textbf{Defining and Registering Symbols.}~~
In order to support complex dependencies, symbol definitions are
completed%
\mnote{MCL: what does \emph{completed} mean here? I think this
  paragraph needs a motivation --- what is it supposed to describe?}
in two different passes. First, symbols are registered in
the \emph{nested} table of another symbol. The
\lstinline!$r$.def($n$,$c$)! method (\cref{lst:def}) queries the
symbol $n$ in $r$, changes the symbol state from undefined to
\texttt{NOT\_READY}, and executes the closure $c$.
\begin{figure}[t]
  \centering
\hrule
\begin{lstlisting}[language=c,morekeywords=function,numbers=left]
  $\mathit{r}$.def($\mathit{name}$, function($\mathit{m}$) {
    @\label{lst:defbegin}@var $\mathit{u}$, $\ldots$; @\Comment{// placeholders for imported symbols}@
    @\label{lst:nested}@$\mathit{m}$.def($\mathit{name'}$, $\ldots$); $\ldots$ @\Comment{// nested symbols}@
    @\label{lst:export}@$\mathit{m}$.exports.$\mathit{k}$ = $\ldots$; $\ldots$ @\Comment{// exported symbols}@
    @\label{lst:basec}@$\mathit{m}$.ctor = $\ldots$; $\mathit{m}$.base = $\ldots$; @\Comment{// (constructor and base, optional)}@
    @\label{lst:mlink}@$\mathit{m}$.mlink = function($\mathit{c}$) { $\mathit{c}$.prototype.$\mathit{m}$ = $\ldots$; $\ldots$ };
    @\label{lst:link}@$\mathit{m}$.link = function() { $\mathit{p}$.prepare(); $\mathit{u}$ = $\mathit{p}$.exported.$\mathit{k}$; $\ldots$ };
  });
\end{lstlisting}
\hrule
  \caption{Structure of a definition closure.}
  \label{fig:def-clos}
\end{figure}
Fig.~\ref{fig:def-clos} presents a schematic view of a symbol
definition, where we provide the structure of the definition closure.
The closure effectively hides all local variable and function names
from outer scopes. It receives the symbol object as parameter $m$ to
fill its definition.
Then, other nested symbols can be defined (\cref{lst:nested}), and 
entries in the export table filled (\cref{lst:export}) to selectively make
inner definitions available (both closures or data). When the symbol
has an associated class, we connect \texttt{ctor} with the class
constructor and, optionally, \texttt{base} with the symbol containing
the base class (\cref{lst:basec}).
The rest of the definition is delayed in other closures, that will be
invoked by \emph{preparing} the symbol (explained below).  Definitions
of class methods, which must be delayed until the constructor is
ready, are delayed in the \texttt{mlink} closure
(\cref{lst:mlink}). On the other hand, values of exported entries of
imported symbols, which are available once the symbol is
\emph{prepared}, are filled in the \texttt{link} closure
(\cref{lst:link}).

\begin{figure}[b]
  \centering
\hrule
\vspace{1em}
\begin{tikzpicture} [
  scale=0.85, every node/.style={scale=0.85},
  font=\sf,
  inbox/.style={draw=Bisque4, text width=3cm, text centered, fill=Bisque1, rounded corners=5pt, inner sep=2pt},
  node distance=1.5em,
  inb/.style={draw, text width=3.0cm, text centered, fill=white, rounded corners=5pt, inner sep=3pt},
  passes/.style={draw=Azure4, text width=3.5cm, text centered, fill=Azure1, inner sep=3pt},
  marrow/.style={->,>=latex},
  darrow/.style={dashed,->,>=latex}
  ]
  \node[passes] (f-e) {Ciao \textbf{Front end}\\ (including expansions)};
  \node[inb] [right of=f-e, xshift=10em] (middle) {Middle level\\(normalized \emph{plain} Prolog)};
  \node[passes] [below of=middle, yshift=-4em] (b-e) {Ciao \textbf{Back end}s \\ -Bytecode \\ -JavaScript};
  \node[inbox] [left of=f-e, yshift=2em, xshift=-12em] (p0) {Input modules\\(\textbf{Prolog/Ciao})};
  \node[inbox] [below of=p0, yshift=-3em] (pkg1) {Packages};
  \draw[marrow] (p0.east) -- (f-e.west);
  \draw[marrow] (pkg1.east) -- (f-e.west);
  \draw[marrow] (f-e) -- (middle);
  \draw[marrow] (middle) -- (b-e);
  \draw[darrow] (p0) -- node[left] {Using} (pkg1);
  \node[inb] [below of=b-e,yshift=-3em,xshift=6em] (jsc) {JS code};
  \node[inb] [below of=b-e,yshift=-3em,xshift=-6em] (other) {Bytecode};
  \draw[marrow] (b-e) edge[->] (jsc);
  \draw[marrow] (b-e) edge[->] (other);
  \draw[darrow] (other) to [in=-90,out=90] node[left,xshift=-1ex,text centered,text width=2cm] {Compilation\\modules} (f-e);
\end{tikzpicture}
\vspace{1em}
\hrule
  \caption{Overview of the Multi Back End Architecture for Ciao.}
  \label{fig:architecture-overview}
\end{figure}
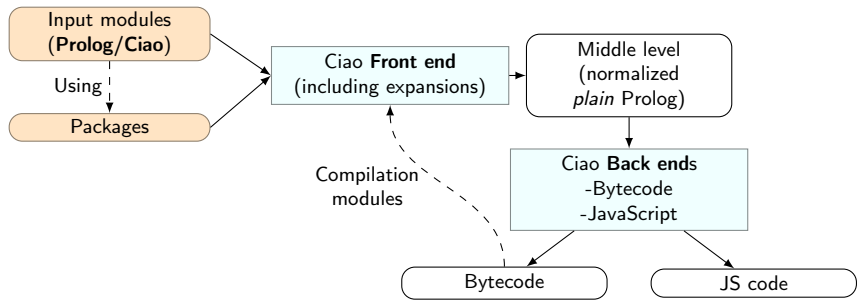

\vspace{0.5ex}
\noindent\textbf{Preparing Symbols.}~~
The \texttt{prepare} method (\cref{fig:modrt-code}-\cref{lst:prepare})
changes the state of \texttt{NOT\_READY} symbols to \texttt{READY}.
First, it prepares the base and fixes the prototype chain of the
constructor (\cref{lst:extends}), and fills the methods invoking
\texttt{mlink}.
Then, values of imported symbols are filled invoking \texttt{link}.
We assume that each \texttt{link} closure calls the \texttt{prepare} 
method of the required symbols. Finally, all nested symbols are
prepared.

\section{Compiler and system architecture for cross-compilation}
\label{sec:compiler}

We base our compiler on two design decisions. First, we share a common
front end with the bytecode back end of Ciao. Second, we reuse most of
the WAM compilation algorithm, and a significant part of the WAM
emulator. A global view of this architecture is shown in
\cref{fig:architecture-overview}. We will elaborate on both points
below.

Sharing the Ciao front end clearly simplifies maintenance of the system,
and avoids undesired or unexpected language incompatibilities. More
importantly, it reuses the Ciao package mechanism for language
extensions. As mentioned before, such packages provide a
collection of syntactic additions (or restrictions) to the input
language, translation rules for code generation to support new
semantics, and the necessary run-time code.
Packages are separated into compile-time and run-time parts. The
compile-time parts (termed \emph{compilation modules}) are invoked
during compilation, and are not necessary during execution. On the
other hand, the run-time parts are only required for execution.
This phase distinction has a number of practical advantages, such as 
reducing executable size. Another important objective achieved is a
stratified separation of modules that makes the code more amenable to
static analysis. In this way, compilation modules are dynamically
loaded by the compiler and invoked during compilation, but not subject
to analysis, and the source modules can be determined statically.
Interestingly, in our context this separation between compile-time
expansion code and run-time code provided by the design of the packages
system also enables cross-compilation without sacrificing
extensibility. Thanks to this, rather than bootstrapping the full system
in JavaScript, we can use the full-fledged compiler, which we have
parameterized to use different back ends as needed during the same
compilation. In this way, compilation modules can be 
compiled and loaded with the bytecode back end, while the source
modules can be compiled independently in the back end selected for
target executables. 

\vspace{0.5ex}
\mnote{MCL: I guess that this para. in the end tries to say that
  optimizations done for the WAM are valid when we're compiling to
  other languages --- I am not sure that the rest is necessary. MH:
  agreed, I simplified a bit. On the other hand only part of the
  translation is WAM based. Added this.}
As mentioned before, our back end is partially based on the WAM.
The combination of a WAM-based engine and
compiler is one of the most efficient approaches to implementing 
Prolog. Such engines and compilers are carefully crafted to optimize
code execution and data 
movements, which makes them relevant and applicable even when the
target of the compilation is a high-level language and the back end a
highly optimized compiler. Although it has been shown that a basic WAM
can be refined from more abstract
specifications~\cite{BoergerRosenzweig90}, the mechanization of such
process is not trivial. For that reason, it is currently unrealistic
to expect that such kinds of optimizations can be introduced
automatically from a high-level compiler.
Thus, our approach reuses
parts of the WAM design and compiler in order to implement relevant
optimization opportunities.

\begin{figure}[t]
  \centering
\small
\hrule
\vspace{1ex}
  \begin{tikzpicture}[
    scale=0.85, every node/.style={scale=0.85},
    font=\sf,
    inbox/.style={draw=Bisque4, text width=5cm, text centered, fill=Bisque1, rounded corners=5pt, inner sep=2pt},
    mbox/.style={draw, text width=5cm, text centered, fill=white, rounded corners=5pt, inner sep=3pt},
    narrow/.style={},
    marrow/.style={->,>=latex}
  ]

\draw[fill=DarkSeaGreen1!50!white, draw=DarkSeaGreen3] (-6,-3.0) rectangle (6,-5.95);

\node[inbox, text width=7cm] (nprolog) at (0,0.1) {
   Normalized \emph{plain} Prolog (no \texttt{;/2}, \texttt{->/2}, $\ldots$)
};
\node[mbox, text width=6cm] (prewam) at (0,-0.5) {
   Pre-WAM (no regs, no mem)    
};

\node[mbox] (wam) at (-3,-1.5) {WAM (X,Y)} ;
\node[mbox] (bytecode) at (-3,-2.5) {Bytecode};
\node[mbox] (emulator)  at (-3,-4.1) {Emulator (C)};
\node[mbox] (runtime)  at (-3,-5.1) {\textbf{Runtime} (C)\\
  Terms, Heap (with GC)\\
  Local frames, Choice points
};

\draw[narrow] (nprolog) -- (prewam); 
\draw[marrow] (prewam) to [in=45,out=225] (wam); 
\draw[marrow] (wam) -- (bytecode) ;
\draw[marrow] (bytecode) -- (emulator) ;
\draw[marrow] (runtime) -- +(0,-1.1) ;

\node[mbox] (jswam) at (3,-1.5) { WAM (args, temp, Y) } ;
\node[mbox] (jscode) at (3,-2.5) { JavaScript Code };
\node[mbox] (jsruntime) at (3,-3.9) {\textbf{Runtime} (JS)\\
Terms (as objects)\\Local frames, Choice points};
\node[mbox] (jslib) at (3,-4.9) {JavaScript Libraries};
\node[mbox] (jsengine) at (3,-5.5) {JavaScript engine (with GC)};

\draw[marrow] (prewam) to [in=135,out=315] (jswam); 
\draw[marrow] (jswam) -- (jscode) ;
\draw[marrow] (jscode) -- (jsruntime) ;
\draw[marrow] (jsengine) -- +(0,-0.7) ;

\draw[fill=DarkSeaGreen3, draw=DarkSeaGreen4] (-6,-6.8) rectangle (6,-6.2);
\node[shape=rectangle,rounded corners=5pt,fill=white] at (0,-6.5) {Operating System \& Libraries};
  \end{tikzpicture}
\vspace{1ex}
\hrule
  \caption{Comparison of the bytecode and JavaScript WAM-based back ends}
  \label{fig:compilation-stack}
\end{figure}
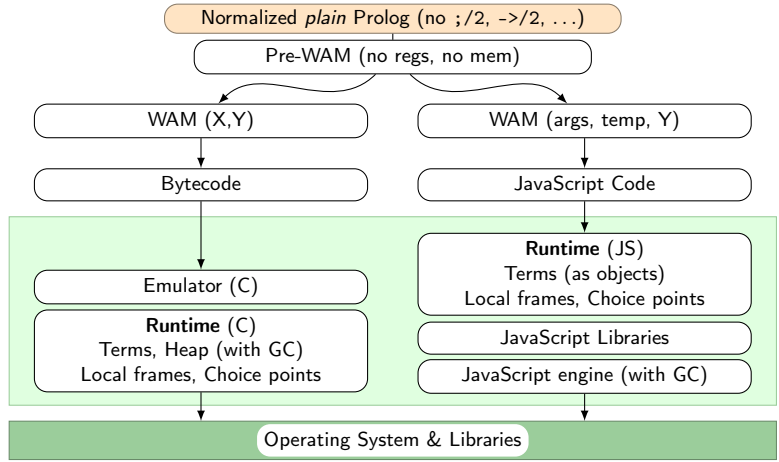

\cref{fig:compilation-stack} shows an overview of our back end,
highlighting the points where it differs from the WAM code generation
performed for the C-based engine. The first step consists of the
normalization of (already expanded) Prolog code into simple Horn
clauses, and the generation of symbolic WAM code. It is at this split
point that %
separate schemes for register assignment, data representation, code
generation, etc.\ are selected, as well as a separate runtime
system. This process is described in the following sections.

\compressection
\subsection{Representing terms and modules}

  Our translation departs from the WAM in that, instead
of defining an explicit heap and using \emph{tagged} words,
we use JavaScript objects to implement terms. \mnote{URGENT: Explain
  how unification is done} Some advantages of this choice (discussed
further in~\cref{sec:experimental}) are that garbage collection is
then performed by the JavaScript engine and that interoperability with
JavaScript code (\cref{sec:foreign}) is simplified.
However, we maintain many of the WAM
concepts within these JavaScript objects. We use subclassing to build a
hierarchy of term constructors. In the following we write $t \le u$ if
$t$ is a subclass of $u$. We define \texttt{term\_base} as the base
class for all terms, and two other base classes for variables
(\texttt{var\_base} $\le$ \texttt{term\_base}), and non-variables
(\texttt{nonvar\_base} $\le$ \texttt{term\_base}).
The \texttt{instanceof} operator can check if a given object belongs
to any particular class, which would already provide the conditional
code necessary to implement all operations on terms, like unification.
However, looking for a particular base class is definitively slower
than fast \emph{switch on tag} operations. In this back end we redefine
most operations on terms using \emph{dynamic dispatching} to simulate
\emph{switch on tag} operations. \cref{fig:def-base} shows an
implementation of these classes. Some methods in this code require a
reference to the state of the \emph{worker} (the set of control stacks
containing trail entries, choicepoints, and frames --- see~\cref{sec:codegen}) denoted as
\texttt{w}.
From that base hierarchy, we define the rest of the elements of the
domain of terms as follows.

\begin{figure}[t]
\compressfigure
  \centering
\hrule
\begin{lstlisting}[language=c,morekeywords=function,numbers=left]
  @\$@m.def("term_base", function(@\$@m) { // Base for terms
    function term_base() {}; @\$@m.ctor = term_base;
  };
  @\$@m.def("var_base", function(@\$@m) { // Base for variables
    function var_base() {}; @\$@m.ctor = var_base; @\$@m.base = @\$@r.query("term_base");
  };
  @\$@m.def("nonvar_base", function(@\$@m) { // Base for non-variables
    function nonvar_base() {}
    @\$@m.ctor = nonvar_base;
    @\$@m.base = @\$@r.query("term_base");
    @\$@m.mlink = function(@\$@c) {
      @\$@c.prototype.unify = function(w, a0) {
        return a0.unify_nonvar(w, this);
      };
      @\$@c.prototype.deref = function() { return this; };
    };
  };
\end{lstlisting}
\hrule
  \caption{Base classes for \emph{var} and \emph{nonvar} terms.}
\compressfigure
  \label{fig:def-base}
\end{figure}

\vspace{0.5ex}
\noindent\textbf{Variables (and Unification)}.~~
Variables are defined in \cref{fig:def-var}.  Since some WAM
optimizations (e.g., conditional trailing) require comparing the
relative age of variables and there is no such order for
JavaScript objects, we define a global timestamp.  Variables are
therefore tuples $(\textit{ref}, \textit{timestamp})$.
Each time a variable is created, the timestamp is
incremented.\footnote{Timestamps on variables are necessary also to
  implement lexical comparisons of terms, such as \texttt{compare/3}.} 
Given a variable $x$, \lstinline!$x$.deref()! \cref{lst:var-deref}
obtains the dereferenced value. It does so by invoking \texttt{deref}
until the variable is unbound. If the variable points to a
\emph{nonvar} object, it simply returns that object, as specified
by the \texttt{deref} of \emph{nonvar}s. Dynamic dispatching is used in a 
similar way to implement unification. The process includes two
cases. First \lstinline!$x$.unify(w, $y$)! (\cref{lst:var-unify})
unifies $x$ and $y$, and returns a boolean indicating whether the
unification succeeded and updates the state \texttt{w} accordingly. The
other case is \lstinline!$x$.unify_nonvar(w, $y$)!
(\cref{lst:var-unify-nonvar}). It assumes that $y$ is dereferenced to
a \emph{nonvar}. For \emph{nonvar} objects, \lstinline!$x$.unify(w, $y$)! is
defined as \lstinline!$y$.unify_nonvar(w, $x$)!, which is implemented
by each of the derived classes.

\begin{figure}[t]
  \centering
\hrule
\begin{lstlisting}[language=c,morekeywords=function,numbers=left]
@\$@m.def("t_var", function(@\$@m) { // Variables
  function v() { this.ref = this; this.timestamp = timestamp++; }
  @\$@m.ctor = v;
  @\$@m.base = @\$@r.query("var_base");
  @\$@m.mlink = function(@\$@c) {
    @\label{lst:var-deref}@@\$@c.prototype.deref = function() { // Dereference
      if (!this.is_unbound()) return this.ref.deref();
      return this;
    };
    @\$@c.prototype.is_unbound = function() { return this.ref === this; };
    @\label{lst:var-unify-nonvar}@@\$@c.prototype.unify_nonvar = function(w, a0) { // Unify with nonvar
      if (!this.is_unbound()) return this.ref.unify_nonvar(w, a0);
      return this.unify(w, a0);
    };
    @\label{lst:var-unify}@@\$@c.prototype.unify = function(w, a0) { // Unify
      if (!this.is_unbound()) return this.ref.unify(w, a0);
      a0 = a0.deref();
      if (a0 instanceof v) {
        if (this.timestamp > a0.timestamp) { 
          this.ref = a0; w.trail(this);
        } else {
          a0.ref = this; w.trail(a0);
        }
      } else { this.ref = a0; w.trail(this); }
      return true;
    };
    @\$@c.prototype.unbind = function() { this.ref = this; }; // Unbind
  };
});
\end{lstlisting}
\hrule
  \caption{Variable definition and methods.}
  \label{fig:def-var}
\end{figure}

\vspace{0.5ex}
\noindent\textbf{Functor, Predicate Symbols, and Modules}.~~They represent
atoms and structures ($\le$\texttt{nonvar\_base}). A constructor
contains as many arguments as its arity, copying them to \texttt{a0},
\texttt{a1}, fields. Additionally it contains static (i.e., shared by
all the objects in the class) entries for the \emph{name} and
\emph{arity}.
Symbols for predicates include an additional \texttt{execute} method
containing the compiled body. The body compilation process will be
described in \cref{sec:codegen}.
During compilation, we generate a new class per predicate or functor
symbol (e.g., \texttt{append/3}, \texttt{./2}, \texttt{[]/0}), nested
within their corresponding module (usually, \texttt{user} for functor
symbols).
Atoms are a special case of functor symbols with arity 0. In the same
way that we associate the body of a predicate with the head functor
definition, we associate the content of a module with its atom. We do
so by storing nested symbols for predicates and functors inside them.
\mnote{add a note about the export table}

\vspace{0.5ex}
\noindent\textbf{Primitive Terms}.~~These are the terms that implement
term wrappers for primitive values or arbitrary JavaScript
objects. They are defined as $t \le$ \texttt{nonvar\_base}. Numbers
(whose class is called \texttt{t\_num}) and  native strings (\texttt{t\_string})
are two examples.
An example definition
for native strings, plugged into the runtime layer, is shown in
\cref{fig:prim-def}. 
\begin{figure}[t]
\compressfigure
  \centering
\small
\hrule
\begin{lstlisting}[language=c,morekeywords=function,numbers=left]
  @\$@r.def("t_string", function(@\$@m) { // String primitive
    function s(a0) { this.a0 = a0; }
    @\$@m.ctor = s;
    @\$@m.base = @\$@r.query("t_nonvar");
    @\$@m.mlink = function(@\$@c) {
      @\$@c.prototype.unbox = function() { return this.a0; }; // Unbox
      @\$@c.prototype.unify_nonvar = function(w, other) { // Unify with nonvar
        if (!(other instanceof s)) return false; // not a string, fail
        return this.a0 === other.a0; // proceed if the strings are the same
      };
    };
  });
\end{lstlisting}
\hrule
  \caption{Primitive term definition for \texttt{t\_string}.}
\compressfigure
  \label{fig:prim-def}
\end{figure}
Code using strings can import the string constructor with \linebreak
\texttt{var s=\$r.query("t\_string").ctor} (only once in its
context). Then, it can be used anywhere with \texttt{new s("...")}.
Note that primitive data creation acts as a \emph{boxing} operation,
while a method \texttt{unbox} is sometimes necessary or convenient.

\compressection
\subsection{Control stacks and code generation}
\label{sec:codegen}

In order to implement backtracking, we adapt some of the registers
and stacks of the WAM, with some modifications. Our machine state is
defined in a \texttt{worker}, that contains:
\begin{itemize}
\item \texttt{goal}: the goal being resolved (a term).
\item \texttt{undo}: a stack that implements the trail. Each entry in
  the trail is a variable. Trailing pushes a variable onto the stack,
  untrailing pops entries up to a certain point, and undoes variable
  changes by invoking the \texttt{unbind} method on each of them.
\item \texttt{choice}: the current choice point, which contains the
  \emph{failure continuation} and a copy of all the worker registers
  (including a reference to \texttt{goal}).
\item \texttt{frame}: the current local frame, which contains Y frame
  variables, saved frame, and the success continuation.
\end{itemize}

The combination of choice, frame, and goal are similar to the frame
structure in B-Prolog~\cite{Cs07aregister-free} (and also to the
original Dec10-Prolog abstract machine): there are no X registers and
argument registers (arguments of \texttt{goal}) and local JavaScript 
variables (for temporaries) are used instead.
The code is generated and executed in a similar way
to~\cite{morales04:p-to-c-padl}. The WAM code is split into chunks
of consecutive instructions separated by predicate calls. Each chunk
is compiled as a closure, which after execution returns the next
closure to be executed.
Before each predicate call, we set a success continuation that points
to the next chunk. When there are no more chunks, we return the next
continuation saved in the worker. As usual, choice points are created
when executing nondeterministic code. Failure is implemented by
untrailing, restoring the worker registers, and jumping to the failure
continuation. The main changes are that the timestamp is also saved
and restored.

\compressection
\compressection
\subsection{Attributed variables}
\label{sec:attvars}

Most current Prolog systems offer (at least some primitive handling
of) {\em attribute variables} in order to be able to extend
unification (and also to allow more flexible control of
execution). Attributed variables, as introduced by
\citeN{Huitouze90:PLILP}, are special variables that can be associated
to a term called an {\em attribute}. Classical built-ins view
attributed variables as normal variables. However the unification of
such a variable with an instantiated term or another attributed
variable is redefined according to a user-defined predicate. This
mechanism is very powerful and allows the efficient
implementation of coroutines~\cite{holzbaur-plilp92}, constraint
solvers~\cite{Holzbaur-clpqr}, and other high-level language
extensions~\cite{HolzbaurFruehwirth99ppdp} directly in Prolog.  To
implement such extensions, a number of Ciao packages %
make extensive use of attributed variables.

In the back end we implement attributed variables obeying the Ciao
interface. They are enabled by the %
\texttt{attr} package, which follows the proposal by
\citeN{demoen02-attributes} for {hProlog}. %
The interface is based on \texttt{get\_attr/3} (that gets the
attribute of a variable), \texttt{put\_attr/3} (that sets the
attribute of a variable), and \linebreak \texttt{attr\_unify\_hook/2} (which is
invoked when two attributed variables are unified, or an attributed
variable is unified with a \emph{nonvar} term).
The runtime code included by the back end for attributed variables
provides a definition for the built-ins, as well as a new class of
terms for attributed variables. 
\mnote{A phrase or two more on the JS implementation would be nice here.}

Once this interface is in place and supported at the JavaScript level
(including a number of additional support predicates) the system can
exploit all the attribute variable-based extensions present in Ciao,
including constraint solvers, extended control, etc. The coroutining 
\texttt{freeze/2} predicate and the Ciao CLP(FD) solver are examples of
such extensions which will be used extensively in the experimental
evaluation.

\compressection
\compressection
\section{Interfacing with JavaScript code}
\label{sec:foreign}

We have focused so far on the runtime and code generation. In
practice, these are useless without a process to make the source and
target layers interoperable. For this reason, we require a foreign
interface between JavaScript and Prolog. This interface is the basis
for both embedding Prolog into existing JavaScript code and
implementing the standard set of libraries interfacing with the
O.S. (in this case, through the browser).

A JavaScript to Prolog interface is straightforward if we follow the
compilation algorithm, which already defines how terms are built and
unified, and how goals are resolved.  Since terms are represented by
JavaScript objects and since memory is reclaimed automatically, there
are no major complications. Only an %
API which abstracts implementation details is necessary.

Accessing JavaScript data and code from the Prolog side is more
involved. Any external object can be seen as an atomic type (so that
two terms bound to JavaScript objects unify iff they are
actually bounded to the same object).  In many cases we want to read or modify
object fields or invoke some of its
methods. In~\cite{DBLP:conf/lpe/WielemakerA02}, a single set of
predicates is used to perform those operations (\texttt{new/2},
\texttt{send/2}, \texttt{get/3}). In our approach, we
follow~\cite{pineda02:ociao} where objects are seen as modules and
methods as predicates of those modules. In practice, this allows using
the same syntax and similar semantics as when specifying
interfaces with formal properties (Ciao-style
\emph{assertions}~\cite{ciaopp-sas03-journal-scp-shortest})
and as in other Ciao foreign interfaces (e.g., for C). 
Consider for example:

\begin{lstlisting}[language=c,morekeywords=function]
:- pred document(-element) + js:foreign("return document;").
:- js:foreign_class element {
  :- pred body(-element) + js:foreign("return this.body;").
  :- pred set_innerHtml(+X) :: string + js:foreign("this.innerHtml=X;").
}.
\end{lstlisting}

\begin{wrapfigure}{r}{0.4\textwidth}
\compressfigure
  \includegraphics[width=0.4\textwidth]{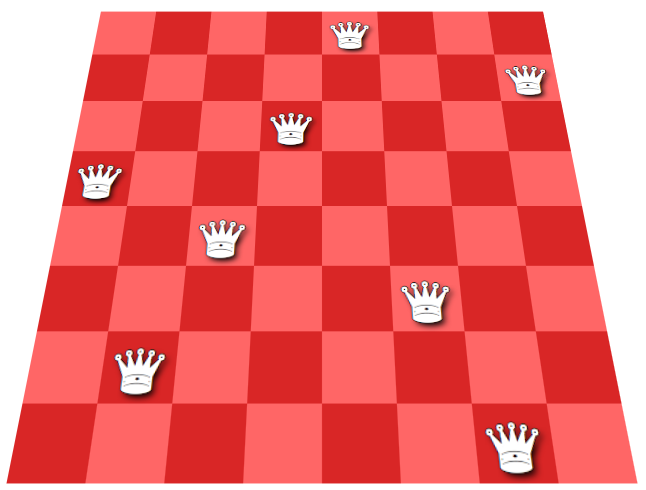}
  \caption{Graphical representation for a solution of Queens-8.}
\compressfigure
  \label{fig:queens8}
\end{wrapfigure}
\noindent
where each \texttt{pred}icate assertion indicates (among other
possible properties) the expected types and modes of the arguments
and the computational properties of the code (\texttt{+} field at end
of the assertion). Foreign code in JavaScript is specified
with the \texttt{js:foreign} property.  Such foreign code is assumed
to be deterministic (Ciao \texttt{is\_det} property) unless otherwise
noted. In the example, modes (\texttt{+} and \texttt{-}) are used in
the usual way to express input and output arguments (Ciao
\texttt{isomodes} library), and types/classes (\texttt{element},
\texttt{string}) are specified in the modes or in a \texttt{::} field.
The assertions inside the \texttt{foreign\_class} block
specify the internal methods associated with objects that are 
\texttt{element}s. Each \texttt{foreign\_class} defines a term wrapper
for foreign data that includes the required glue code predicates.
Given the previous interface, the following is a simple, 
html-oriented \emph{hello world} program: 
\begin{verbatim}
main :- document(D), D:body(B), B:set_innerHtml("Hello World").
\end{verbatim}
It queries the document body and replaces its text with the given
string. Using the same idea %
we have easily created more complex code like that generating
\cref{fig:queens8}.\footnote{The on-line version of this program 
is available at 
\url{http://cliplab.org/~jfran/ptojs/queens_ui/queens_ui.html}
and can be tested on, e.g., a smart phone (a QR code is provided for
convenience).} 
\compressection
\section{Experimental evaluation}
\label{sec:experimental}

We have measured experimentally the performance of the compiler
back end and the system runtime and libraries by compiling a collection
of unmodified, small and medium-sized benchmarks to JavaScript and
comparing their execution time (under the V8 engine and Chrome
17~\cite{google:v8/chrome}) with that of the Ciao virtual machine.
Although raw performance is currently not our main goal, 
this gives us an initial indication of the %
size of problem that is amenable to client-side execution with the
current implementation. 
We have chosen a) the following classical benchmarks:

\newcommand{\mdes}[1]{\parbox{7em}{\textbf{#1}}}

\begin{description}
\item[\mdes{qsort}] Implementation of QuickSort.
\item[\mdes{tak}] Computation of the Takeuchi function.
\item[\mdes{fft}] Fast Fourier transform.
\item[\mdes{primes}] Sieve of Eratosthenes.
\item[\mdes{nreverse}] Naive reversal of a list using \texttt{append/3}.
\item[\mdes{deriv}] Symbolic derivation of polynomials.
\item[\mdes{poly}]  Raises symbolically the expression
  \texttt{1+x+y+z} to the n$^{\mathrm{th}}$ power.
\item[\mdes{boyer}] Simplified Boyer-Moore theorem prover kernel.
\item[\mdes{crypt}] Cryptoarithmetic puzzle involving multiplication.
\item[\mdes{guardians}] Prison guards playing game.
\item[\mdes{jugs}] Jugs problem.
\item[\mdes{knights}] Chess knight tour, visiting only once every board cell.
\item[\mdes{11-queens}] $N$-Queens with $N = 11$.
\item[\mdes{query}] \parbox[t]{0.77\linewidth}{Makes a natural language
    query to a knowledge database with in\-for\-ma\-tion about country
    names, population, and area.}\smallskip
\end{description}

\noindent
as well as b) the following collection of more complex problems:

\begin{description}
\item[A collection of CLP(FD) programs.] We use a
  CLP(FD) library based on indexicals~\cite{clpfd} written 
  in Prolog \emph{using attributed variables} (plus syntactic extensions,
  etc.).  We tested the classical \textbf{\texttt{SEND+MORE=MONEY}}, a
  \textbf{\texttt{sudoku}} solver, the \textbf{\texttt{bridge}}
  optimization problem, and the first solution to \textbf{$N$-Queens}
  with $N = 50$.
\item[\mdes{sat-freeze}] A benchmark based on an implementation of the
  DPLL algorithm for solving the Boolean satisfiability problem
  (SAT)~\cite{DBLP:conf/flops/HoweK10}. The solver implements
  \emph{watched literals} using \texttt{freeze/2} for delayed control.
\end{description}

\begin{figure}[t]
\compressfigure
  \centering
  \begin{tikzpicture}
\pgfplotsset{
    compat=newest,
    width=5.5cm,
    title style = {yshift = 10pt,name=title},
    xtick align = inside,
    xtick pos = both,
    xticklabel pos = upper,
    yticklabel style={name=yticklabel},
    xmin=-5,
    xmax = 65,
    ytick align = outside,
    ytick pos = left,
    yticklabel pos=left,
    ytick=data,
    y dir = reverse,
    xbar,
    nodes near coords,
    every node near coord/.append style = {anchor=west},
    yticklabel shift=0.25cm
}

\matrix[column sep=4pt, row sep=2pt]{
\begin{axis}[
    height=4cm,
    yticklabel pos=left,
    yticklabels from table={tables/data1.dat}{name},
    bar width=4pt,
    ymin = 0,
    ymax = 5
]

\addplot[fill=red!50!white] table [
y=Y-Position,
x=Slowdown,
] {tables/data1.dat};
\end{axis}
&
\begin{axis}[
    height=4cm,
    yticklabel pos=right,
    yticklabels from table={tables/data2.dat}{name},
    ytick pos = right,
    bar width=4pt,
    ymin = 0,
    ymax = 5
]

\addplot[fill=red!50!white] table [
y=Y-Position,
x=Slowdown,
] {tables/data2.dat};
\end{axis}
\\
\begin{axis}[
    height=4.5cm,
    yticklabel pos=left,
    yticklabels from table={tables/data3.dat}{name},
    bar width=4pt,
    xticklabels = none,
    ymin = 0,
    ymax = 7
]

\addplot[fill=red!50!white] table [
y=Y-Position,
x=Slowdown,
] {tables/data3.dat};
\end{axis}
&
\begin{axis}[
    height=4.5cm,
    yticklabel pos=right,
    yticklabels from table={tables/data4.dat}{name},
    ytick pos = right,
    bar width=4pt,
    xticklabels = none,
    ymin = 0,
    ymax = 6
]

\addplot[fill=red!50!white] table [
y=Y-Position,
x=Slowdown,
] {tables/data4.dat};
\end{axis}

\\
};
\end{tikzpicture}
  \caption{Slowdown comparison of the different groups of benchmarks.}
  \label{fig:comparison-speed-plot}
\compressfigure
\end{figure}
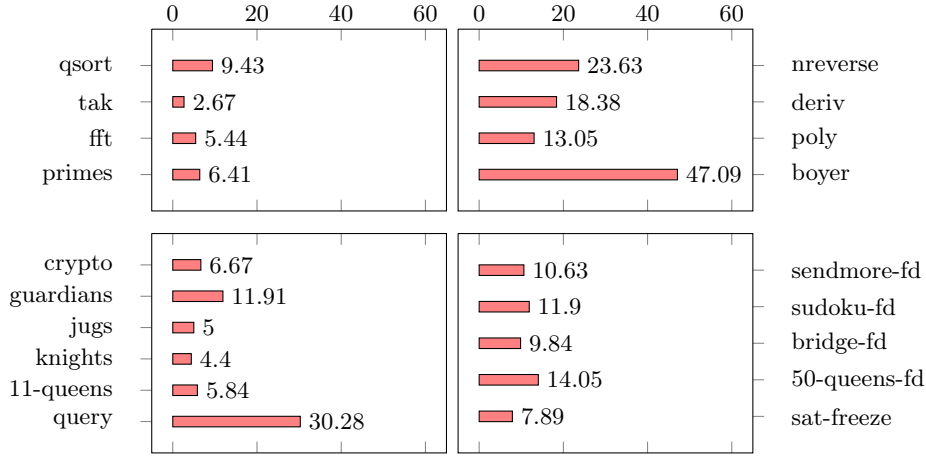

We ran %
on a MacBook Pro, Intel Core 2 Duo (2.66 GHz and
3MB L2 cache). The execution times and the slowdown ratios are shown in
\cref{fig:comparison-speed-plot} and \cref{tab:comparison-speed}.  

\begin{table}[t]
\begin{tabular}{llrrr}
\hline
& \textbf{Benchmark}  & \textbf{Ciao} $t_0$ (ms) & \textbf{JS (V8)} $t_1$ (ms) & \textbf{Ratio} $t_1/t_0$ \\
\hline
\emph{Group 1}
& \textbf{qsort (x1000)} & 28.39 & 267.00 & 9.43 \\
& \textbf{tak (x10)} & 55.70 & 149.00 & 2.67 \\
& \textbf{fft} & 13.60 & 74.00 & 5.44 \\
& \textbf{primes (x100)} & 4.21 & 27.00 & 6.41 \\
\hline
\emph{Group 2}
& \textbf{crypt (x10)} & 6.60 & 44.00 & 6.67 \\
& \textbf{guardians} & 6.63 & 79.00 & 11.91 \\
& \textbf{jugs (x10)} & 17.80 & 89.00 & 5.00 \\
& \textbf{knights} & 371.00 & 1636.00 & 4.40 \\
& \textbf{11-queens} & 286.00 & 1672.00 & 5.84 \\
& \textbf{query (x100)} & 17.30 & 524.00 & 30.28 \\
\hline
\emph{Group 3}
& \textbf{nreverse} & 4.02 & 95.00 & 23.63 \\
& \textbf{deriv (x1000)} & 6.42 & 118.00 & 18.38 \\
& \textbf{poly (x10)} & 22.50 & 295.00 & 13.05 \\
& \textbf{boyer} & 27.20 & 1281.00 & 47.09 \\
\hline
\emph{Group 4}
& \textbf{sendmore-fd} & 20.40 & 217.00 & 10.63 \\
& \textbf{sudoku-fd} & 110.00 & 1310.00 & 11.90 \\
& \textbf{bridge-fd} & 2322.00 & 22857.00 & 9.84 \\
& \textbf{50-queens-fd} & 151.00 & 2123.00 & 14.05 \\
& \textbf{sat-freeze (x10)} & 376.00 & 2969.00 & 7.89 \\
\hline
\end{tabular}
  \caption{Performance comparison of the bytecode and JavaScript back ends.}
  \label{tab:comparison-speed}
  \vspace{-1.1ex}
\end{table}

Since the target is a dynamic language where we do not have precise
control of data sizes, placement, memory movements, or assembler
instructions (unlike, e.g., in a translation to C or a bytecode engine
written in C), the gap between the %
source code and what is finally executed is large and slowdowns are to
be expected.  Indeed, the geometric mean of the slowdown for all the
benchmarks is 10.00.
Also, in our experience the actual performance is highly dependent on
the engine which executes the JavaScript code (see later for details).

A more careful study splits the benchmarks into several groups. The
first group (from \texttt{qsort} to \texttt{primes}) requires fast
management of control and backtracking, but does not create complex
data, and exhibits the best performance results. The second group
(from \texttt{nreverse} to \texttt{boyer}) heavily depends on data
creation and unification and performs worse. In particular, we tracked
down this difference to a concrete issue in the state of our
compilation scheme: we currently perform less indexing than what
the WAM can do.
To validate this assumption, we disabled indexing in the
\textbf{bytecode} version of \texttt{boyer}. This yielded code which
is 7.68 slower, which made the JavaScript / bytecode speed ratio to be
in the ballpark of the first group. 
The performance of the benchmarks in the third (search problems) and
fourth groups (constraints) can be explained in a similar way by their
internal dependency on complex data manipulation (e.g., internal data 
structures for FD implementation) or indexing (e.g., \texttt{query}).

\vspace{0.5ex}
\noindent\textbf{Practicality.}~~
We believe that the absolute performance achieved is sufficient for a
large range of interactive, web-bound, non computationally-intensive
tasks.
And even for the case of more computationally-intensive tasks, if they 
are, however, not straighforward to program in a traditional language
(e.g., they involve constraint solving, reasoning, etc.) we believe
that our technology can be very useful. As just a simple example,
rule-based form validations with  non-trivial, interdependent
rules can be complex enough to be much more natural to program in a
constraint/logic-based language. When confronted with the dilemma
of server-side vs.\ client-side execution, data transmission
delays\footnote{Note that the lower bounds to latency times
  (\texttt{ping}) are limited by the speed of light transmission over
  optical lines:  6.7 ms / 1000 km, and, in fact, much
  higher in practice.}  may make client-side execution be preferable 
even if client-side execution is slower than on the
server side. And, as mentioned before, there are also other issues
such as privacy or the cost of server-side computation and storage, 
which favor 
client-side execution irrespective of performance. 

As stated before, our first goal has been the construction of a
framework that is as complete as possible and easy to maintain, and
which offers a high degree of compatibility for existing code.  We
believe our results show that our system allows non-trivial code (such
as that implementing CLP(FD) or the SAT solver example, just to give
two examples of code foundations which can be interesting to execute
in a browser) to be easily run %
in browsers practically unmodified, alongside with other JavaScript
code.
Nevertheless, %
we expect to obtain further performance
improvements %
using more sophisticated compilation techniques 
(e.g., applying indexing and other WAM optimizations more aggressively
and eventually 
program analysis). At the same time, the approach will obviously
benefit from future improvements in the JavaScript platforms (which we
have seen to improve significantly during our work).

\vspace{0.5ex}
\noindent\textbf{Further Details on the Performance Results.}~~
One source of overhead identified is the cost of term
representation. 32-bit WAM implementations typically require only
$(1+n)$ words to represent an $f/n$ structure constructor, and $1$
word for simpler objects like variables and constants.
Objects in JavaScript are in principle much more complex, as values
are records (dictionaries or hash tables) with an arbitrary number of
fields.  One key optimization in V8 are \emph{hidden classes} and
\emph{inline caching} (brought in from efficient implementations of
Self~\cite{Chambers:1989:EIS:74878.74884}) to represent records
efficiently and optimize property access.  Even with those
optimizations the system still requires 3 words plus data per object.

In~\cite{tagschemes-ppdp08} we observed that simply doubling the
space required for tagged words, leaving the rest of the engine
unaltered, can significantly affect performance in WAM-based 
machines. 
Marking some distinguished elements in the type lattice with tag bits
is a clever optimization, hard to reproduce without a specialized heap
representation.
Moreover, we need additional data such as \emph{timestamps} that is
not required in the WAM, where the age of terms can be compared
directly by their pointer addresses.

Alternatively, an explicit management of the heap as an array could
improve term encoding (by making it closer to that of the WAM).
This approach has been used in the compilation from C to JavaScript (as
done in EMScripten~\cite{Zakai:2011:ELC:2048147.2048224}, which has a 2.4-8.4 slowdown w.r.t. native code)
or systems compiling to Java like Lean
Prolog~\cite{Tarau:2011:IST:2076022.1993497}.  However, by adopting
this approach we would lose some advantages of using a native
JavaScript representation, including getting garbage collection for
free.  Additionally, larger and more complex runtime code would be
required. The impact of this change is difficult to evaluate \emph{a
  priori} and is left as future work.

\vspace{0.5ex}
\noindent\textbf{Performance on Other Browsers.}~~
We tested the performance of our compilation on other major browsers
(SpiderMonkey, Firefox 11; Nitro, Safari 5.1.2), and observed a
slowdown of between
2.1$\div$ and 11.5$\div$ w.r.t.\ Google's V8.
We have verified in synthetic benchmarks that, despite these virtual
machines being as good as V8 for usual JavaScript programs (including
a handcoded version of \texttt{tak}), they are less efficient in the
creation and manipulation of many small objects.  One reason is that
objects in V8 are significantly smaller than in other engines,
which makes SpiderMonkey and Nitro less convenient for our current
translation scheme.  We conjecture that V8 implements optimizations 
related to frequent, small object creation, as one of the benchmarks in the
V8 suite is the compilation of the EarleyBoyer classic Scheme
benchmarks using \texttt{sch2js}, which later evolved to be part of 
Hop~\cite{tfp2007:hop-client}.
Nevertheless, for benchmarks which do not create large numbers of
objects, the engines of the major browsers offer similar performance.
It would also be interesting to explore whether with an explicit heap 
smaller performance gaps might be observed across engines.

\compressection
\section{Conclusions and future work}
\label{sec:conclusions}

We believe our system makes a significant contribution towards the
practical feasibility of client-side Web applications based (fully or
partially) on (constraint) logic programming, while relying
exclusively on Web standards. This reliance makes it possible to
execute code on a variety of devices without any need for installation
of additional plug-ins or proprietary code. We believe this is an
important advantage, specially since a good number of the currently
popular portable devices make such installation hard or impossible.

We made a strong effort to preserve source compatibility with existing
Prolog code, and declaring special libraries and dialectic changes
explicitly. For all this, the module and package system of Ciao was of
great help. 
The current implementation represents a promising scaffolding on top
of which a truly full-fledged system can be built. Our future work 
will be focused on several parallel lines. First, developing automatic
methods for distributing code across browsers and servers, using AJAX
or WebSockets for communication. Second, improving the compilation
technology, specially using more WAM-level optimizations, analysis
information, and the combination of such optimizations with JIT
compilation, which has already been shown to significantly improve the
execution of Prolog interpreters~\cite{BoLeSch2010}.
Third, gradually extending the current implementation of the Ciao 
libraries and language features, which would allow client-side
execution of more and more complex programs.

The system is integrated in the Ciao repository and will be included
in upcoming Ciao distributions.  Examples and benchmark programs
(including the Queens program of \ref{fig:queens8}) are publicly
available from \url{http://cliplab.org/~jfran/ptojs}.

\bibliographystyle{acmtrans}
\bibliography{../../../bibtex/clip/clip,../../../bibtex/clip/general}

\end{document}